\documentclass{jpconf}
\usepackage{amssymb}
\usepackage{amsmath}
\usepackage[T1]{fontenc}
\usepackage[english]{babel}
\usepackage{graphicx}
\begin{document}
\title{Holographic dark energy at the Ricci scale}
\author{Iv\'{a}n Dur\'{a}n  and  Diego Pav\'on}
\address{Departament de F\'isica, Universitat Aut\`onoma de
Barcelona, 08193 Bellaterra, Spain} \ead{ ivan.duran@uab.cat,
diego.pavon@uab.es}
\begin{abstract}
We consider a holographic cosmological model in which the infrared
cutoff is fixed by the Ricci's length and dark matter and dark
energy do not evolve separately but interact non-gravitationally
with one another. This substantially alleviates the cosmic
coincidence problem as the ratio between both components remains
finite throughout the expansion. We constrain the model with
observational data from supernovae, cosmic background radiation,
baryon acoustic oscillations, gas mass fraction in galaxy
clusters, the history of the Hubble function, and the growth
function. The model shows consistency with observation.
\end{abstract}

\section{Introduction}\label{introduction}
Holographic models of late acceleration have become of fashion,
among other things, because they link the dark energy density  to
the cosmic horizon, a global property of the universe, and have a
close relationship to the spacetime foam \cite{jng1},
\cite{arzano}. For a summary motivation of holographic dark
energy, see section 3 in \cite{cqg-wd}. The cosmological model we
present here assumes a spatially flat homogeneous and isotropic
universe dominated by dark matter (DM) and dark energy (DE)
(subscripts $M$ and $X$, respectively), the latter obeying the
holographic relationship
\begin{equation}
 \rho_{X}= \frac{3 M_{P}^{2}\,  c^{2}}{L^{2}}\, ,
\label{rhox}
\end{equation}
where $c^{2}$ is a dimensionless parameter that we will take as
constant -though strictly speaking it may slowly vary with time
\cite{jcap-nd}-, and $L = (\dot{H} +2H^{2})^{-1/2}$ denotes the
Ricci's lenght. We adopt the latter because it corresponds to the
size of the maximal perturbation leading to the formation of a
black hole \cite{brustein}. For models that identify $L$ with
Hubble's length, $H^{-1}$, see \cite{jcap-idw} and references
therein.

\begin{figure}[htb]
\begin{center}
\includegraphics[width=10cm]{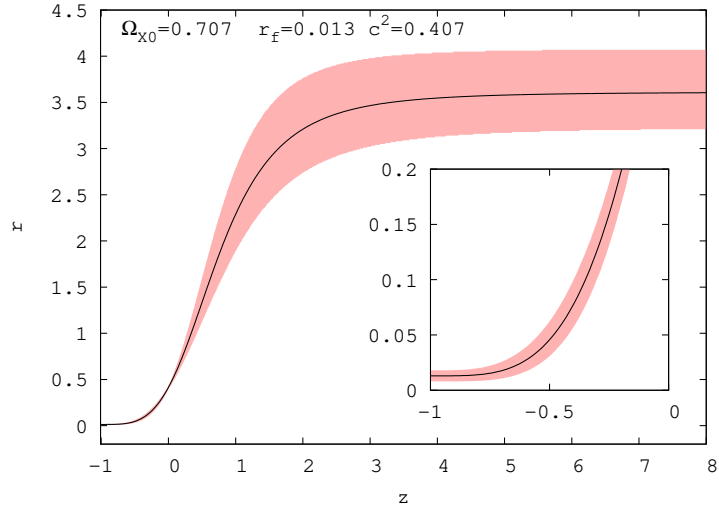}
\end{center}
    \caption{\label{fig:r} Plot of the ratio $r$ between the energy densities vs. redshift
    for the best-fit model. As the inset shows, $r_{f} \equiv r(z)$  does
    not vanish when $z \rightarrow -1$. In this, and the next figure, the red
    swath indicates the region obtained by including the $1\sigma$ uncertainties of
    the constrained parameters used in the calculation.}
\end{figure}

\begin{figure}[htb]
 \hspace{-0.5 cm}
\begin{minipage}{0.3\textwidth}
\centering
\includegraphics[width=8cm]{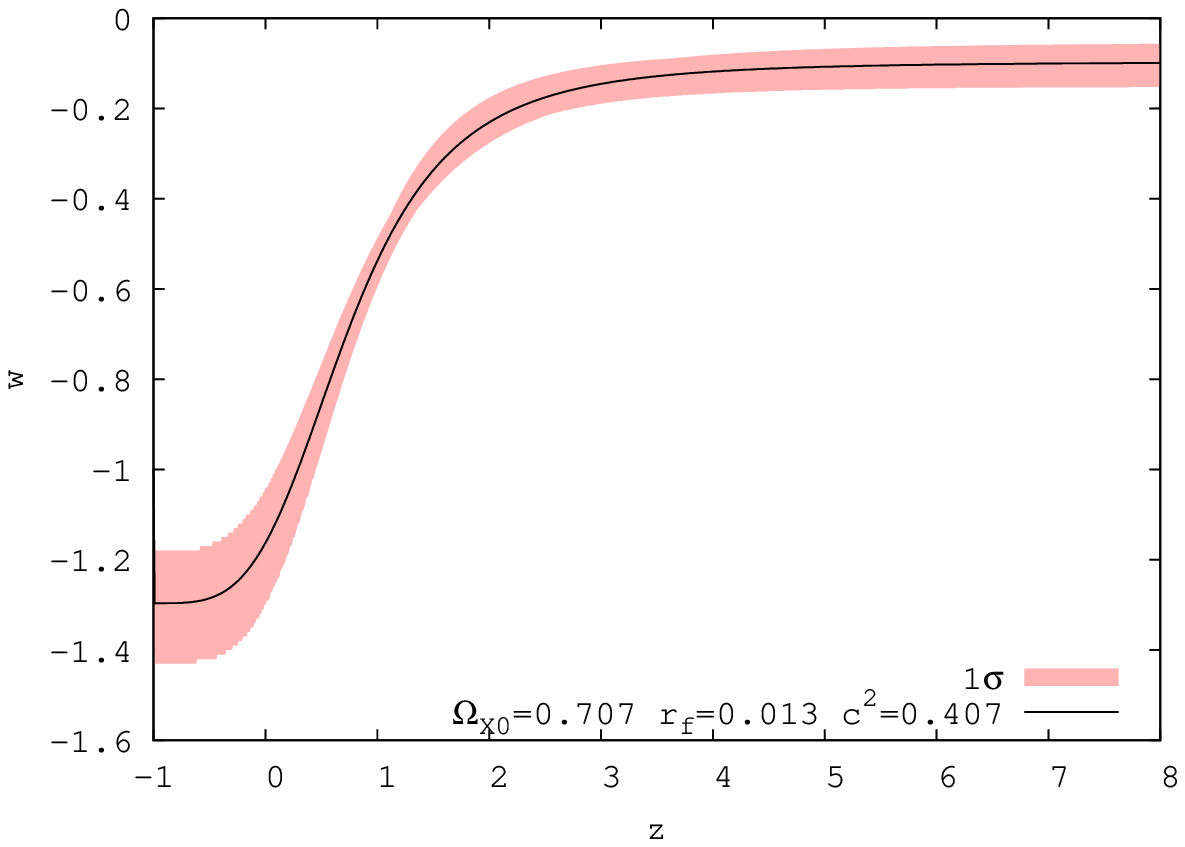}
\end{minipage}
 \hspace{3.5 cm}
\begin{minipage}{0.3\textwidth}
\centering
\includegraphics[width=8cm]{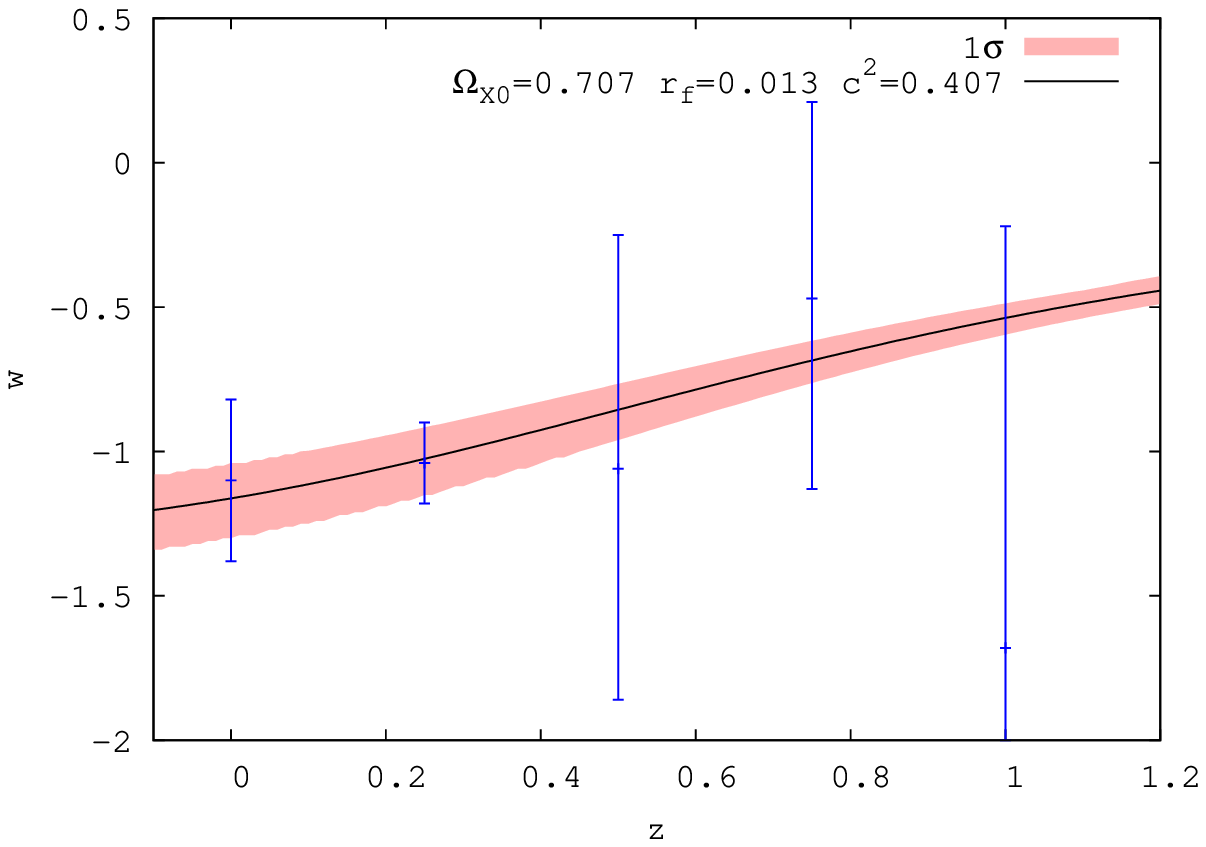}
\end{minipage}
\caption{The equation of state parameter  vs. redshift up to $z =
8$ (left panel), and up to $ z = 1.2$ only (right panel) for the
best fit holographic model. At high redshifts $w$ approaches the
equation of state of non-relativistic matter and at low redshifts
it does not depart significantly from $-1$. The observational data
are taken from  \cite{serra-prd}. The error bars indicate a
$2\sigma$ uncertainty.} \label{fig:w}
\end{figure}

The other main assumption is that DE and DM  interact
non-gravitationally with each other according to
\begin{eqnarray}\label{eq:EvolIn}
\dot{\Omega}_{M} -  \left(1-\frac{2\Omega_{X}}{c^{2}}\right)(1-\Omega_{X})\, H&=&QH \, ,\\
\label{eq:EvolIn2} \dot{\Omega}_{X}+
\left(1-\frac{2\Omega_{X}}{c^{2}}\right)(1-\Omega_{X})\, H&=&-QH
\, .
\end{eqnarray}
Here $\Omega_{M}$ and $\Omega_{X}$ stand for the fractional
densities of the components, and the interaction term
\begin{equation}
 Q = -\frac{r_{f}}{(1+r_{f})^{2}}\left(1+r_{f}-\frac{2}{c^{2}}\right)
\, ,
\label{eq:QChim}
\end{equation}
is such that ratio $r \equiv \Omega_{M}/\Omega_{X}$ evolves from a
constant value at early times to a final, finite, value -denoted
as $r_{f}$- at late times (see \cite{ivan-diego} for details and
Fig. \ref{fig:r}). This clearly alleviates the coincidence problem
(namely, ``why are the densities of matter and dark energy of the
same order precisely today?" \cite{steinhardt}), something beyond
the reach of the $\Lambda$CDM model. A consequence of the model is
that the equation of state of dark energy $w \equiv
p_{X}/\rho_{X}$ varies with expansion as shown in Fig.
\ref{fig:w}.

\section{Contrasting the model with observation}\label{contrasting}
The model has four free parameters ($\Omega_{X0}$, $c^{2}$,
$r_{f}$, and $H_{0}$) which we constrained with observational data
from supernovae (557 data points \cite{Amanullah}), cosmic
background radiation \cite{komatsu}, baryon acoustic oscillations
(at $z = 0.35$ \cite{Eisenstein} and $z = 0.2$ \cite{Percival}),
gas mass fraction in galaxy clusters (42 data points from x-ray
measurements \cite{Allen}), the Hubble function history (15 data
points \cite{Riess-2009}, \cite{gazta}, \cite{simon}, and
\cite{stern}), and the growth function (5 data points from Table 2
in \cite{Gong}). (See \cite{ivan-diego} for details).

Data from the growth function -which correspond to matter
perturbations within the horizon- are particularly relevant
because they can more easily discriminate between cosmological
models with a similar history of the Hubble function. For the
model under consideration the growth function of matter, $f \equiv
d \ln \delta_{M}/d \ln a$, is governed by
\begin{equation}\label{eq:growthFactor}
f'+f^{2}+\left(\frac{1}{c^{2}(1+r)}+Q \frac{1+r}{r}\right)f-
\frac{3r^{3}+2Q^{2}(1+r)^{3}-2Q r(2+r-r^{2})}{2r^{2}(1+r)}=0 \, .
\end{equation}
In the non-interacting limit, $Q \rightarrow 0$, this expression
reduces to the corresponding one of the Einstein-de Sitter
cosmology, namely: $f' + f^{2} + [2 \, + \, (\dot{H}/H^{2})] f =
3\Omega_{M}/2$.

\begin{figure}[htb]
 \hspace{-0.5 cm}
\begin{minipage}{0.3\textwidth}
\centering
\includegraphics[width=5.5cm]{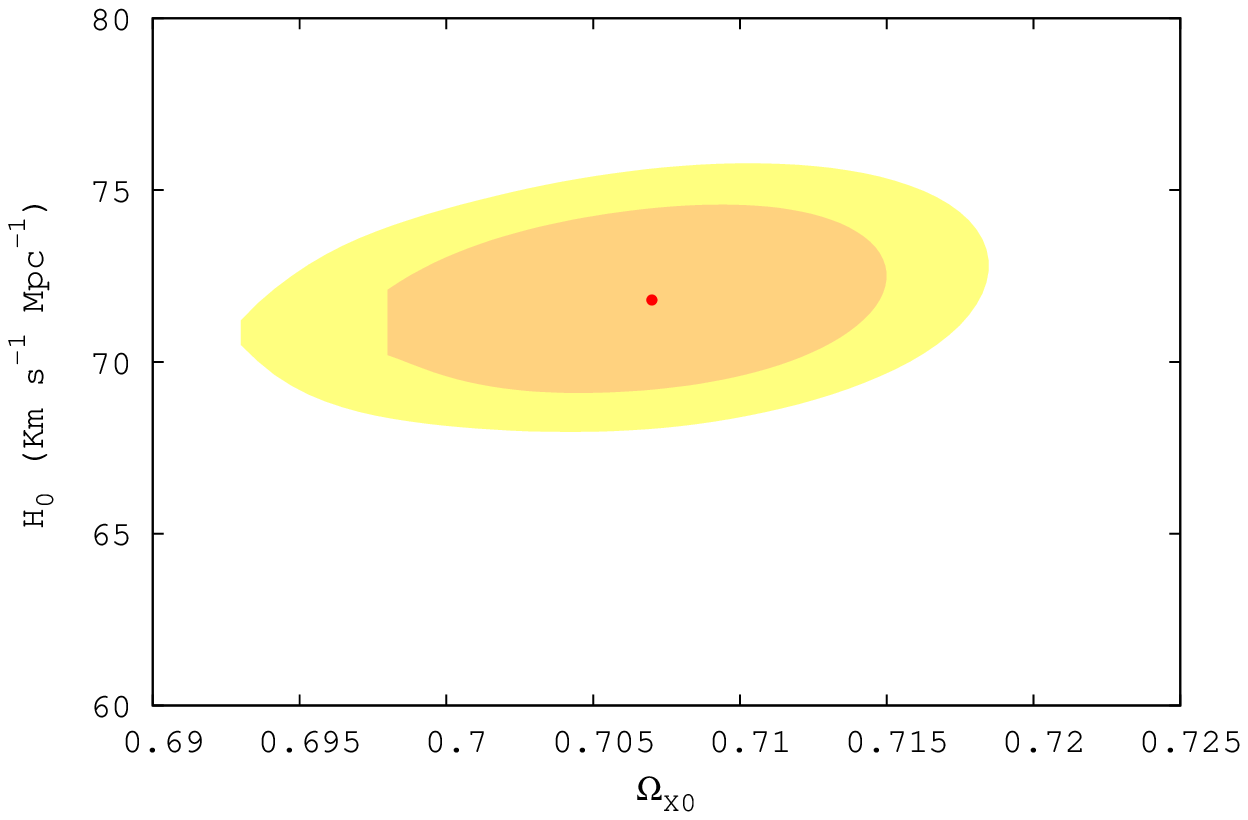}
\end{minipage}
 \hspace{1 cm}
 \begin{minipage}{0.3\textwidth}
\centering
\includegraphics[width=5.5cm]{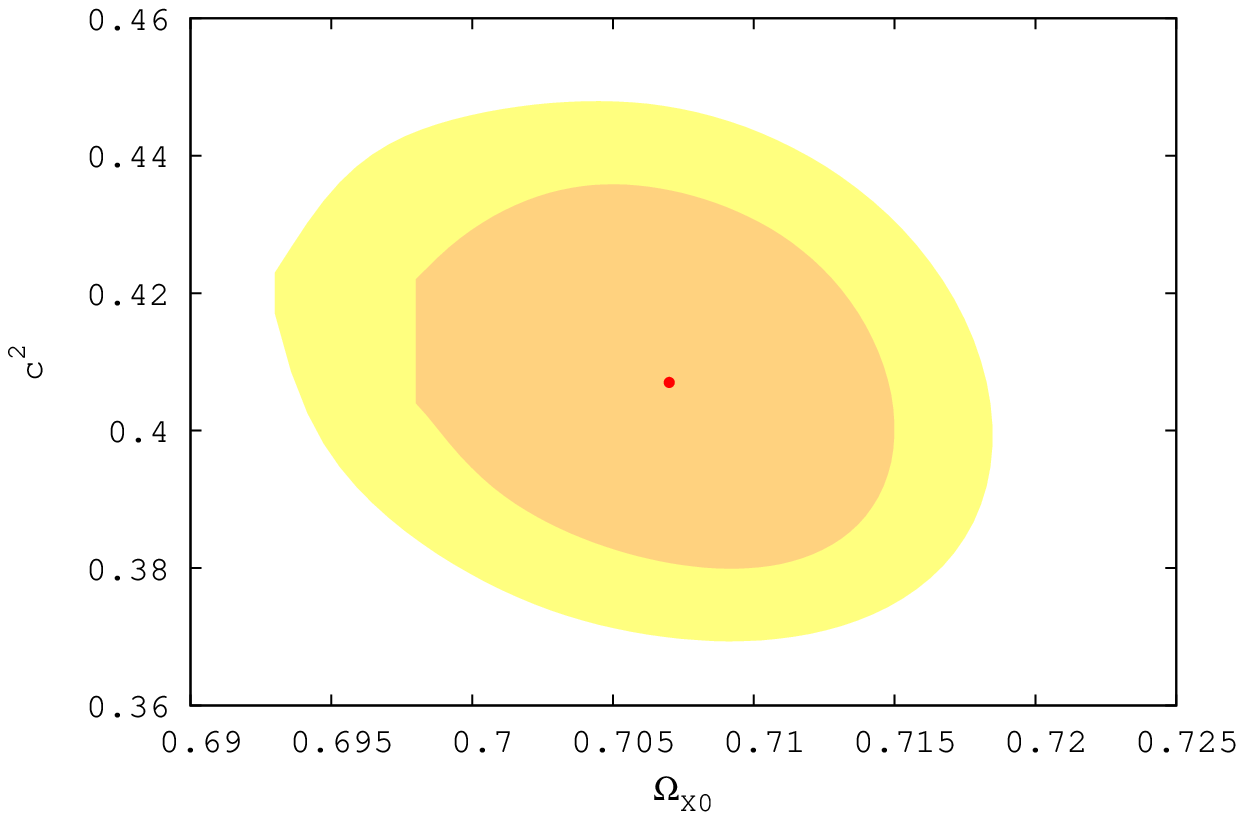}
\end{minipage}
\hspace{1 cm}
\begin{minipage}{0.3\textwidth}
\centering
\includegraphics[width=5.5cm]{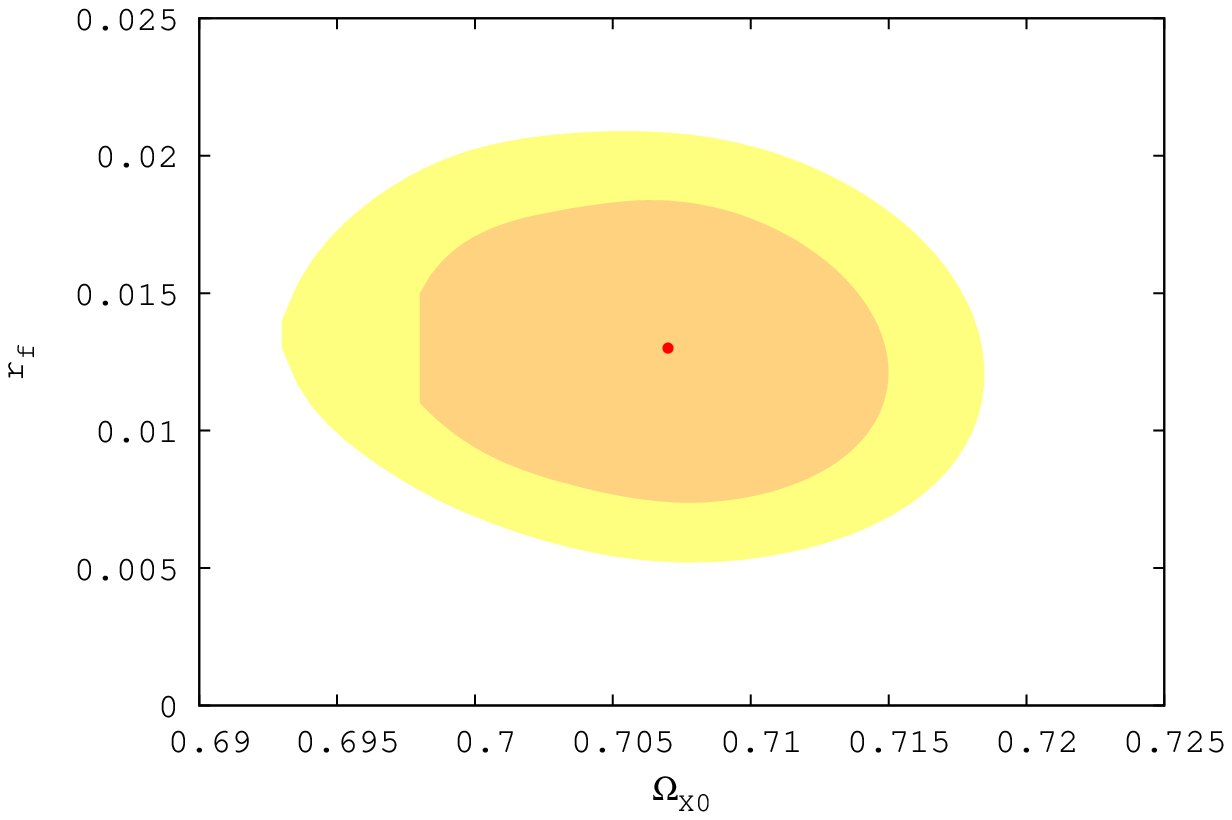}
\end{minipage}
\caption{Panels from left to right show the 68.3\% and 95.4\%
confidence contours for the pairs of free parameters
($\Omega_{X0}$, $H_{0}$), ($\Omega_{X0}$, $c^{2}$),
($\Omega_{X0}$, $r_{f}$), respectively, obtained by constraining
the holographic model with SN Ia+CMB-shift+ BAO+x-ray+H(z)+growth
function data. The solid point in each panel locates the best fit
values.} \label{fig:3elipses}
\end{figure}

\begin{figure}[htb]
 \hspace{-0.5 cm}
\begin{minipage}{0.3\textwidth}
\centering
\includegraphics[width=8cm]{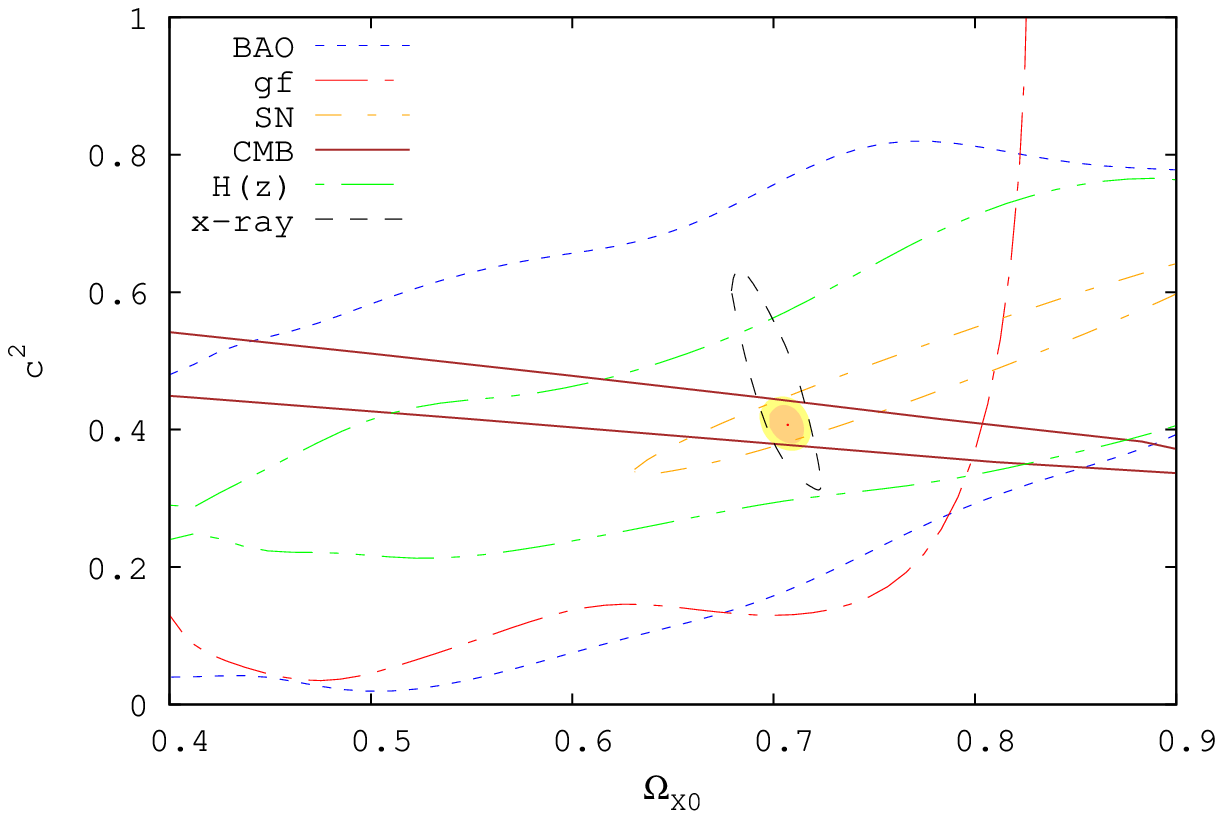}
\end{minipage}
 \hspace{3.5 cm}
\begin{minipage}{0.3\textwidth}
\centering
\includegraphics[width=8cm]{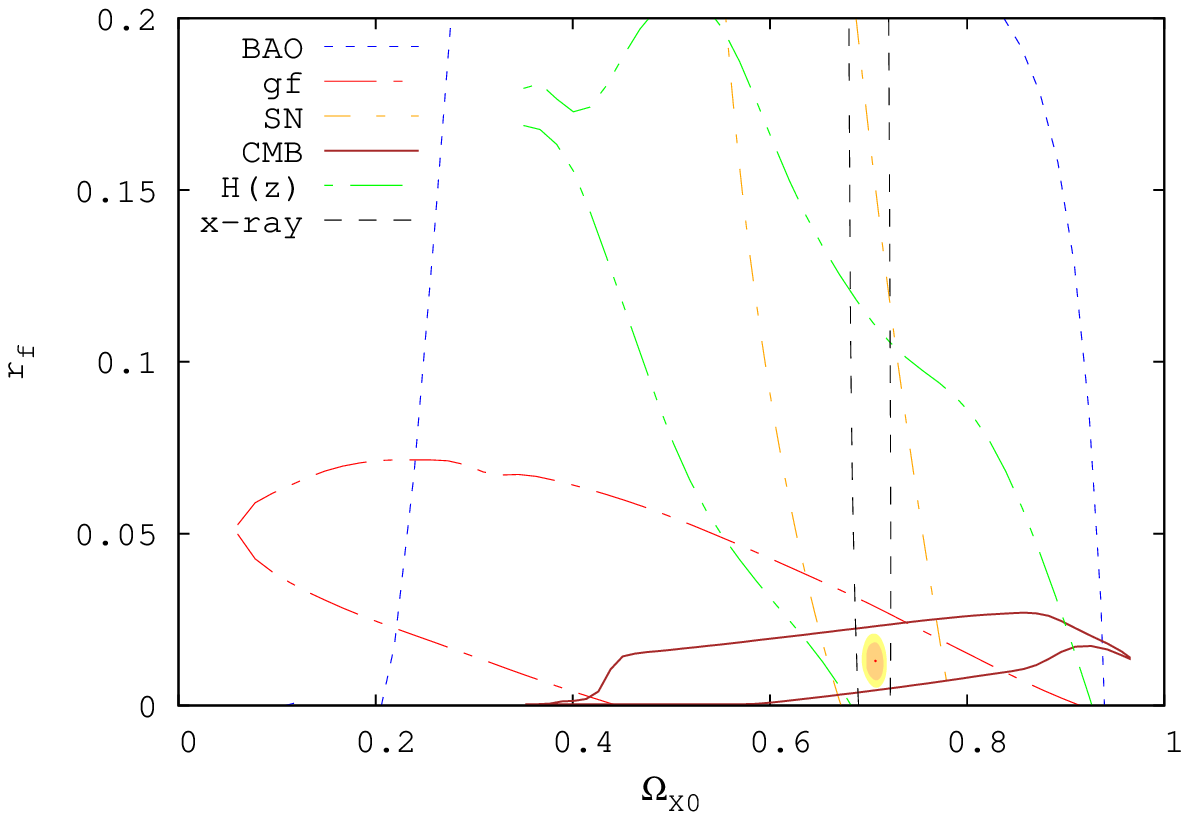}
\end{minipage}
\caption{Left panel: Probability contours for SN Ia, CMB, BAO,
x-ray, $H(z)$, and growth function,
    in the ($\Omega_{X0}, c^{2}$) plane. The joined constraint uses $ \chi^{2}_{\rm total} =
    \chi^{2}_{sn} \, + \, \chi^{2}_{cmb}\, + \, \chi^{2}_{bao} \, + \, \chi^{2}_{x-rays}+ \,
    \chi^{2}_{Hubble}\, + \, \chi^{2}_{gf}$. Right panel: Idem
    for the ($\Omega_{X0}, r_{f}$) plane.} \label{fig:2elipses}
\end{figure}

The best-fit values of the free parameters are found to be:
$\Omega_{X0}=0.707\pm0.009$, $c^{2}=0.407^{+0.033}_{-0.028}$,
$r_{f}= 0.013^{+0.006}_{-0.005}$, and $H_{0}=71.8\pm2.9$ km/s/Mpc.
This together Table \ref{table:chi2} and  figures
\ref{fig:3elipses} and \ref{fig:2elipses} summarizes our findings.
It is worthy of note that the non-interacting case, $Q = 0$ (which
implies $r_{f} = 0$ via Eq. (\ref{eq:QChim})), lies over $2\sigma$
away from the best fit value. This seems to be  a generic feature
of holographic models.

Table \ref{table:chi2} shows the partial, total, and total
$\chi^{2}$ over the number of degrees of freedom of the
holographic model along with the corresponding values for the
$\Lambda$CDM model. The latter  has just two free parameters,
$\Omega_{M0}$ and $H_{0}$. Their best-fit values after
constraining the model to the same sets of data are $\Omega_{M0} =
0.266 \pm 0.006$, and $H_{0} = 71.8 \pm 1.9 \, $km/s/Mpc.

\begin{table}
\caption{\label{table:chi2} $\chi^{2}$ values of the best-fit
holographic model and the best fit $\Lambda$CDM model.}
\begin{tabular}{ p{2.1 cm} p{1.6 cm} p{1.2 cm} p{1.2 cm} p{1.4 cm}p{1.4 cm}p{1.4 cm} p{1.5 cm}
p{1.3 cm}} \br
Model & $\chi^{2}_{sn}$ & $\chi^{2}_{cmb}$ & $\chi^{2}_{bao}$ &  $\chi^{2}_{x-rays}$ &  $\chi^{2}_{H}$ & $\chi^{2}_{gf}$ & $\chi^{2}_{{\rm total}}$ & $\chi^{2}_{{\rm total}}/dof$  \\
\hline
Holographic  & $543.70$ & $0.01$ & $1.20$ & $41.79$ & $9.57$ & 1.06 & $597.34$ & $\; \; 0.96$ \\
\hline
$\Lambda$CDM & $542.87$ & $0.05$ & $1.13$ & $41.59$ & $8.73$ & 0.43 & $594.80$ & $\; \; 0.96$ \\
\br
\end{tabular}
\end{table}
\section{Conclusions}
The statistical analysis sketched above shows that the ratio
$\chi^{2}_{total}/dof$ is lower than unity for the holographic
interacting model of section \ref{introduction}, whereby it is
compatible with observation. However, as Table \ref{table:chi2}
shows, the $\Lambda$CDM model fits better the same sets of data.
Yet, the latter cannot explain the cosmic coincidence problem
while the former can.

\ack{ID was funded by the ``Universidad Aut\'{o}noma de Barcelona"
through a PIF fellowship. This research was partly supported by
the Spanish Ministry of Science and Innovation under Grant
FIS2009-13370-C02-01, and the ``Direcci\'{o} de Recerca de la
Generalitat" under Grant 2009SGR-00164.}



\section*{References}
\begin{thebibliography}{99}
\bibitem{jng1} Ng YJ 2001 {\it Phys. Rev. Lett.} \textbf{86} 2946
\bibitem{arzano} Arzano M, Kephart TW and Ng YJ 2007
{\it Phys. Lett.} B \textbf{649} 243
\bibitem{cqg-wd} Zimdahl W and Pav\'{o}n D 2007 {\it Class. Quantum
Grav.} \textbf{24} 5461
\bibitem{jcap-nd} Radicella N and Pav\'{o}n D 2010 {\it
J. Cosmol. Astropart. Phys.} JCAP10(2010)005
\bibitem{brustein}
Brustein R 2008 ``Cosmological Entropy Bounds" in {\it String
Theories and Fundamental Interactions} (Lecture Notes in Physics
vol 737) eds M Gasperini and J Maharana (Heidelberg: Springer) pp
619-659
\bibitem{jcap-idw} Dur\'{a}n I, Pav\'{o}n D and Zimdahl W 2010 {\it
J. Cosmol. Astropart. Phys.} JCAP07(2010)018
\bibitem{ivan-diego} Dur\'{a}n I and Pav\'{o}n D 2011 {\it Phys.
Rev.} D, in the press ({\it Preprint} arXiv:1012.2986
[astro-ph.CO])
\bibitem{steinhardt}
Steinhardt PJ 1997 ``Cosmological Challenges for the 21st Century"
in {\it Critical Problems in Physics} eds VL Fitch {\it at al}
(Princeton: Princeton University Press) pp 123-146
\bibitem{serra-prd}
Serra P {\it et al} 2009 {\it Phys. Rev.} D  \textbf{80} 121302
\bibitem{Amanullah} Amanullah R {\it et al} 2010 {\it Astrophys. J.}
(in the press) {\it Preprint} arXiv:1004.1711
\bibitem{komatsu} Komatsu E {\it et al} 2009 {\it Astrophys.
J. Suppl.} \textbf{180} 330
\bibitem {Eisenstein} Eisenstein DJ {\it et al} 2005
{\it Astrophys. J.} \textbf{633} 560
\bibitem {Percival} Percival WJ \textit{et al} 2010 {\it Mon. Not. R. Astron.
Soc.} \textbf{401} 2148
\bibitem{Allen} Allen SW {\it et al} 2008 {\it Mon. Not. R. Astron.
Soc.} \textbf{383} 879
\bibitem{Riess-2009} Riess AG {\it  et al} 2009 {\it Astrophys.
J.} \textbf{699} 539
\bibitem{gazta} Gazta\~{n}aga E, Cabr\'{e} A, and Hui L 2009
{\it Mon. Not. R. Astron. Soc.} \textbf{399} 1663
\bibitem{simon} Simon J, Verde L and Jim\'{e}nez R 2005 {\it Phys.
Rev.} D \textbf{71} 123001
\bibitem{stern} Stern D, Jim\'{e}nez R, Verde L,
Kamionkowski M and Stanford SA 2010 {\it J. Cosmol. Astropart.
Phys.} JCAP02(2010)008
\bibitem{Gong} Gong Y 2008 {\it Phys. Rev.} D \textbf{78} 123010
\end{thebibliography}
\end{document}